\documentclass[]{aa}
\usepackage{graphics,latexsym,epsfig,graphicx}     
\begin{document}
\title{Cosmic shear as a tool for precision cosmology: minimising intrinsic galaxy alignment-lensing interference} 
\author{Lindsay J. King} 
\institute{Institute of Astronomy, University of Cambridge, Madingley Rd, Cambridge CB3 0HA\\ }
\date{} \authorrunning{Lindsay King} 
\titlerunning{Decoupling intrinsic-lensing interference} 
\abstract{Cosmic shear leads to a correlation of the
observed ellipticities of galaxies, an effect which is used to place constraints on cosmological parameters, and to explore the evolution of dark matter and dark energy in the universe. However,  a possible systematic 
contaminant of the lensing signal is intrinsic galaxy alignment, with a correlation length of a few Mpc.
Hirata \& Seljak (2004) have recently demonstrated that
for some models of intrinsic distortions, there may also be a cross-correlation between 
the intrinsic and lensing signals, which may dominate the intrinsic signal, and suppress the lensing power spectrum by several tens of percent. Unlike the pure intrinsic signal, this new term cannot be
accounted for by neglecting or down-weighting pairs of galaxies which are physically close.
Extending the correlation function tomography method of King \& Schneider (2003) we illustrate how 
the impact of both intrinsic and cross-correlations can be significantly reduced, in the context of surveys with photometric redshift information. For a ground-based cosmic shear survey of $\sim 100$ sq. degrees with photometric redshifts, even in the presence of systematic contaminants at the level considered here, cosmological models degenerate in the $\Omega_{\rm m}-\sigma_8$ plane can be distinguished well in excess of the 3-$\sigma$ level.
} 
\def\A{{\cal A}}
\def\eck#1{\left\lbrack #1 \right\rbrack}
\def\eckk#1{\bigl[ #1 \bigr]}
\def\rund#1{\left( #1 \right)}
\def\abs#1{\left\vert #1 \right\vert}
\def\wave#1{\left\lbrace #1 \right\rbrace}
\def\ave#1{\left\langle #1 \right\rangle}
\def\arcsecf {\hbox{$.\!\!^{\prime\prime}$}}
\def\arcminf {\hbox{$.\!\!^{\prime}$}}
\def\bet#1{\left\vert #1 \right\vert}
\def\vp{\varphi}
\def\vt{{\vartheta}}
\def\map{{$M_{\rm ap}$}}
\def\d{{\rm d}}
\def\mj{$\rm {m_{j}}$}
\def\mk{$\rm {m_{k}}$}
\def\col{$\rm {m_{j}}-\rm {m_{k}}$}\def\eps{{\epsilon}}
\def\vc{\vec} 
\def\s{{\rm d}}
\def\s{{\rm s}}
\def\t{{\rm t}}
\def\E{{\rm E}}
\def\L{{\cal L}}
\def\i{{\rm i}}
\def\seps{{\sigma_{\epsilon}}}
{\catcode`\@=11
\gdef\SchlangeUnter#1#2{\lower2pt\vbox{\baselineskip 0pt \lineskip0pt
  \ialign{$\m@th#1\hfil##\hfil$\crcr#2\crcr\sim\crcr}}}}
\def\gtrsim{\mathrel{\mathpalette\SchlangeUnter>}}
\def\lesssim{\mathrel{\mathpalette\SchlangeUnter<}}      
\maketitle
\section{Introduction}
The tidal gravitational field of mass inhomogeneities distorts the
images of distant galaxies, resulting in correlations in their
observed ellipticities. This cosmic shear signal depends upon cosmological 
parameters and the matter power spectrum (Blandford et al. 1991; 
Miralda-Escud\'e 1991; Kaiser 1992). 
In 2000, four teams announced the first detections of cosmic
shear (Bacon et al. 2000; Kaiser et al. 2000; van Waerbeke et
al. 2000; Wittman et al. 2000; Maoli et al. 2001).
The current and next generation of multi-colour surveys will cover 
hundreds or thousands of square degrees, and have the potential to 
measure the dark matter power spectrum at small scales (inaccessible 
to CMB measurements) with unprecedented precision, and furthermore 
to probe its evolution. These  surveys will tighten errors on 
cosmological parameters, particularly when combined with results from the 
CMB, SNIa and galaxy surveys (e.g. van Waerbeke et al. 2002). 
For example, Tereno et al. (2005) forecast the errors on parameters from 
CFHTLS (Canada-France-Hawaii Telescope Legacy Survey) coupled with CMB, primarily from WMAP (Wilkinson Microwave Anisotropy Probe), noting that
many combinations benefit greatly from the joint constraint.   
   
A possible systematic contaminant
of the lensing correlation function ($\xi^{\rm L}$) is intrinsic 
alignment, which may arise during the galaxy formation process. This has been 
subject to numerical, analytic and 
observational studies [e.g. Croft \& Metzler 2000; Heavens et al. 2000 (HRH); 
Crittenden et al. 2001; Catelan et al. 2001; Mackey et al. 2002; Brown et
al. 2002; Jing 2002; Hui \& Zhang 2002]. However, a great deal of uncertainty remains, with 
amplitude estimates spanning more than an order of magnitude.
 
Recently, Hirata \& Seljak (2004; hereafter HS04) demonstrated that there may be a finite 
intrinsic alignment-lensing cross-term ($\xi^{\rm LI}$), depending on the mechanism giving 
rise to intrinsic alignments. One such mechanism is tidal stretching: the shape of a galaxy is determined by the shape of a halo in 
which it forms - which is in turn related to the tidal field where the halo
resides. If this tidal field also lenses a background galaxy, then one might expect a correlation between the ellipticities of the foreground and background galaxies. HS04 estimated that the cross-term can potentially suppress the lensing power spectrum by several tens of percent, adopting an intrinsic alignment model where the mean ellipticities of galaxies are linear functions of the tidal field in which they form. However, they note that this cross-term is strongly model dependent - for instance it vanishes for a quadratic rather than a linear dependence on the tidal field.

It has been shown that photometric
redshift information could be used to suppress $\xi^{\rm I}$, 
by downweighting or ignoring galaxy pairs at approximately the same
redshift (King \& Schneider 2002), or by downweighting nearby
pairs and subtracting a model of the
intrinsic alignment signal from the observed ellipticity correlation
function (Heymans \& Heavens 2003). Since any cross-term does not arise from physically close galaxies, such schemes could not deal with this systematic. In King \& Schneider (2003), hereafter 
KS03, a correlation function tomography method was developed to isolate the intrinsic and lensing-induced components of the galaxy ellipticity correlation function, assuming that photometric redshift 
information is available, so that the correlation function can be measured 
between different redshift slices. 

In this paper, using the fact that the cross-term signal has a different dependence on redshift than either the pure lensing or intrinsic signals, being negligible when galaxies in a pair have a small separation in redshift, we show that a similar procedure significantly reduces any systematic suppression of the extracted lensing signal, and corresponding error in the lensing analysis. 

\section{General method and specific assumptions}
In this Section, the method proposed by 
KS03 is summarised, and extended to account for the possibility of a cross-term between intrinsic alignment and lensing.

\subsection{Method\label{gen}}
The E-mode ellipticity correlation function for galaxies with angular separation 
$\theta$, and at true redshifts $z_{i}$, $z_{j}$, is composed of a 
lensing signal, an intrinsic signal, and a cross-term
\begin{equation}
\xi(\theta,z_{i},z_{j}) = \xi^{\rm L}(\theta,z_{i},z_{j}) + \xi^{\rm I}(\theta,z_{i},z_{j})\ + \xi^{\rm LI}(\theta,z_{i},z_{j})\,.
\end{equation}
The origin and behaviour of $\xi^{\rm I}$, and of $\xi^{\rm LI}$ is not yet well understood. 
However, the physics behind the pure lensing contribution is well established,  $\xi^{\rm L}(\theta,z_{i},z_{j})$ being related to the 3-dimensional matter power spectrum $P_\delta$ through
\begin{eqnarray}
&&\xi^{\rm L}(\theta,z_i,z_j)=
{9 H_0^4 \Omega_{\rm m}^2\over 4 c^4}
\int_0^{{\rm min}[w_{i},w_{j}]} {\d w\over a^2(w)}~~~~~~\nonumber \\
&&\times R(w,w_{i})\,R(w,w_{j}) 
 \int{\d \ell\,\ell\over (2\pi)}\,P_\delta\rund{{ \ell\over f(w)},w}\,{\rm
J}_0(\ell\theta) \,,
\label{powspec}
\end{eqnarray}
where $H_0$ and $\Omega_{\rm m}$ are the values of the Hubble
parameter and matter density parameter at the present epoch, and $a(w)$ is
the scale factor at comoving distance $w$, normalised such that
$a(0)=1$ today. J$_{0}$ is the $0$-th
order Bessel function of the first kind, and $\vec\ell$ is the angular
wave-vector. The comoving distance at $z_{i}$ is denoted by 
$w_{i}$, and $f(w)$ is the comoving angular diameter distance, which depends on the
spatial curvature $K$:
\begin{equation}
  f(w) = \left\{
  \begin{array}{ll}
    K^{-1/2}\sin(K^{1/2}w) & (K>0)\\
    w & (K=0)\\
    (-K)^{-1/2}\sinh[(-K)^{1/2}w] & (K<0) \\
  \end{array}\right\}\;.
\end{equation}
The function $R(w,w')=f(w'-w)/f(w')$ is the ratio
of the angular diameter distance of a source at comoving distance $w'$
seen from a distance $w$, to that seen from $w=0$. 

When we account for the availability of photometric redshift
estimates rather than spectroscopic ones, the galaxy ellipticity 
correlation function becomes  
\begin{equation}
{\bar \xi}(\theta,{\bar z_{i}},{\bar z_{j}})=\int{\rm d}z_{i}\int{\rm
d}z_{j}\,p(z_{i},z_{j}|{\bar z_{i}},{\bar z_{j}},\theta)\,\xi(\theta,z_{i},z_{j})\;,
\label{phot}
\end{equation}
where $p(z_{i},z_{j}|{\bar z_{i}},{\bar z_{j}},\theta)$ is 
the probability to have true redshifts $z_i$ and $z_j$ given
photometric estimates ${\bar z_{i}}$ and ${\bar z_{j}}$. This is given by
\begin{eqnarray}
&&p(z_{i},z_{j}|{\bar z_{i}},{\bar z_{j}},\theta)=\\\nonumber
&&\frac{p(z_{i}|{\bar z_{i}})\,p(z_{j}|{\bar z_{j}})\,[1+\xi_{\rm gg}(r)]}{\int{\rm d}z_{i}\int{\rm d}z_{j}\,p(z_{i}|{\bar z_{i}})\,p(z_{j}|{\bar z_{j}})\,[1+\xi_{\rm gg}(r)]}\;,
\end{eqnarray}
where $\xi_{\rm gg}$ is the galaxy spatial correlation function, which
may also include redshift dependent evolution, and $r$ is the comoving 
separation of galaxies in a pair. 

One can determine 
${\bar \xi}(\theta,{\bar z_{i}},{\bar z_{j}})$ on a 3-dimensional grid 
of $N_{K}$ angular separation bins of width $\Delta\theta$ centred on 
$\theta_{K}$ (index $K$), and $N_{Z}$ photometric redshift bins of width $\Delta z$ centred on each of
${\bar z_{i}}$ (index $I$) and ${\bar z_{j}}$ (index $J$). 
We will refer to an {\em observed} signal using the notation ${\bar \xi}^{\rm obs}_{IJK}$, and to a
theoretical prediction for a particular cosmology, intrinsic alignment 
mechanism, and cross-term using $\ave{{\bar \xi}^{\rm mod}}_{IJK}$. 

The lensing, intrinsic and cross-term correlations can be expressed in terms of
sets of template functions $A_{n}$, $B_{n}$ and $C_{n}$
\begin{eqnarray}
&&\bar {\xi^{\rm L}}(\theta,{\bar z_{i}},{\bar z_{j}})=\sum_{n=1}^{{\rm N_{L}}}a_{n}A_{n}(\theta,{\bar z_{i}},{\bar z_{j}})\;,\\\nonumber
&&\bar {\xi^{\rm I}}(\theta,{\bar z_{i}},{\bar z_{j}})=\sum_{n=1}^{{\rm N_{I}}}b_{n}B_{n}(\theta,{\bar z_{i}},{\bar z_{j}})\;,\\\nonumber
&&\bar {\xi^{\rm LI}}(\theta,{\bar z_{i}},{\bar z_{j}})=\sum_{n=1}^{{\rm N_{LI}}}c_{n}C_{n}(\theta,{\bar z_{i}},{\bar z_{j}})\;,
\end{eqnarray}
where $a_{n}$, $b_{n}$ and $c_{n}$ are the amplitudes of the $n$-th
lensing, intrinsic and cross-term template functions respectively. As noted in KS03, extra 
functions can be added as required, to span the range of plausible models. We describe our
choice of models in Sect.\,\ref{basis} below. 
A single index $m$ identifies correlations between
two bins with redshift indices $I$ and $J$, and with angular separation index $K$.
There are $N_{M}=N_{Z}\,(N_{Z}+1)\,N_{K}/2$ such independent
measurements. The total of $N=N_{L}+N_{I}+N_{LI}$ gridded template models for
the correlation functions
can be written as an $N_{M}\times N$ design matrix ${\cal M}$, and their amplitudes as an $N$-dimensional column vector ${\cal G}$ so that
\begin{equation}
\ave{{\bar\xi}^{\rm mod}}_{m}={\cal M}_{mn}{\cal G}_{n}\;.
\end{equation}
We recover $\bar{\xi}^{\rm obs}_{m}$ in terms of the template
functions: using the method of least squares, the best-fit 
estimates of the amplitudes ${\cal G}_{n}$ are those values ${\hat {\cal G}}_{n}$ 
which minimise
\begin{equation}
S=(\vec{\bar{\xi}}^{\rm obs}-{\cal M}{\cal G})\,{\cal C}^{-1}\,(\vec{\bar{\xi}}^{\rm
obs}-{\cal M}{\cal G})\;,
\end{equation} 
where ${\cal C}$ is the covariance matrix. Since $\ave{\bar{\xi}^{\rm mod}}_{m}$
is a linear combination of template functions, the linear least squares 
estimators are 
\begin{equation}
{\hat {\cal G}}=\left({\cal M}^{T}{\cal C}^{-1}{\cal
M}\right)^{-1}{\cal M}^{T}{\cal C}^{-1}\vec{\bar{\xi}}^{\rm obs}\;.
\label{rec}
\end{equation}
With an observed correlation function $\vec{\bar{\xi}}^{\rm obs}$, a design matrix ${\cal M}$ containing our choice of template functions, and ${\cal C}$, we can obtain ${\hat {\cal G}}$ i.e. the projection of $\vec{\bar{\xi}}^{\rm obs}$ into the template functions. This means that separate fits for the lensing, intrinsic and cross contributions can be obtained. Most importantly, for the purpose of using the lensing signal in a subsequent analysis, is that we reduce the impact of any contamination from the intrinsic alignment-lensing cross-term as well as from the instrinsic alignment.

\subsection{Covariance matrix\label{spec}}
To evaluate (\ref{rec}), the covariance matrix ${\cal C}_{mm'}$ is required. Following KS03 a simplified model which neglects the cosmic variance contribution will be used; thus the elements of ${\cal C}$ come from the intrinsic ellipticity dispersion of the source galaxies. This is the dominant contribution to the covariance at small angular scales; at larger angular scales, the cosmic variance terms start to dominate, with the transition angular scale depending on the survey geometry (Kaiser 1998; Schneider et al.\ 2002). 

The elements of the covariance matrix are 
\begin{equation}
{\cal C}_{m\,m'}=
\frac
{2\left(\seps^{2}/2\right)^{2}}
{{\cal N_{\rm p}}(m)}\left[
\delta_{m,m'}
\left(1\,+\,\delta_{I,I'}\delta_{J,J'}\delta_{I,J}
\right)
\right]\;,
\end{equation}
where there is an extra contribution from auto-variance terms. Since ${\cal C}$ is diagonal, 
the elements of its inverse are simply 
$\left({\cal C}^{-1}\right)_{m\,m'}=\delta_{m\,m'}/{\cal C}_{m\,m}$.
The galaxy ellipticity dispersion is denoted by $\seps$. In bin $m$,
the number of pairs is given by
\begin{equation}
{\cal N_{\rm p}}(m)={\cal N_{\rm F}}\,\left[n_{0}p(z_{I})\Delta
z\right]\,\left[n_{0}p(z_{J})\Delta z\right]L^{4}\frac{\Delta\theta}{L}\tau\left(\frac{\theta}{L}\right)\;,
\end{equation} 
where $n_{0}$ is the galaxy 
number density, $p(z_{I})$ is the redshift probability density for redshift 
bin $I$, and $L$ is the extent of the field, assumed to be square. 
$\tau(\theta/L)$ is a function that takes into account the fact that fields 
have finite extent, and is evaluated numerically as in KS03.

\subsection{Basis models\label{basis}}
The simple, restricted, set of template functions used here is described below. 
\subsubsection{Lensing basis models}
The lensing template functions $A_{n}$ are the gridded
${\bar \xi^{\rm L}}(\theta,{\bar z_{i}},{\bar z_{j}})$ for 3 models of
the underlying cosmology: i) $\Lambda$CDM ($\Omega_{\rm m}=0.3$, cosmological constant density parameter $\Omega_{\Lambda}=0.7$), ii) OCDM ($\Omega_{\rm m}=0.3$, $\Omega_{\Lambda}=0$) and iii) $\sigma$CDM
($\Omega_{\rm m}=1$, $\Omega_{\Lambda}=0$); for simplicity, 
the power spectrum normalisation $\sigma_{8}=0.9$, and shape parameter $\Gamma=0.21$ for each model. We could choose templates close to $\Lambda$CDM, as described in Section 4.

Starting with a 3-dimensional primordial power spectrum $P(k)\propto k$, the Bardeen et al. (1986) transfer function is used to describe its evolution, along with the prescription of 
Peacock \& Dodds (1996) for evolution in the non-linear regime. 
The lensing correlation functions are calculated using the relationship between the power spectrum
and $\xi^{\rm L}$ given in (\ref{powspec}), and then integrated over the
photometric redshift uncertainties as in (\ref{phot}).
Here it is assumed that $p(z|{\bar z})$ is a Gaussian with dispersion
$\sigma_{\rm phot}$, centred on ${\bar z}$. 

\subsubsection{Intrinsic alignment basis models}
Nine template models $B_{n}$ for the intrinsic alignments are
considered, the true spatial intrinsic correlation function being 
parameterised in terms of a correlation length $R_{\rm corr}$, and an exponent $\alpha$:   
\begin{equation}
\eta(r,z)=(1+z_{\rm av})^{\alpha}\left[{\rm exp}\,\left(-r/R_{\rm corr}\right)\right]\;,
\end{equation}
where $z_{\rm av}$ is the mean redshift of galaxies in a
pair and $r$ is their comoving separation. We use the approximation 
$r^{2}=(w_{i}-w_{j})^{2}\,+\,\theta^{2}f^{2}\left[(w_{i}+w_{j})/2\right]$.
$R_{\rm corr}$ is taken to be [1, 3, 10]$\,h^{-1}\,{\rm Mpc}$
and $\alpha$ to be [$-$1, 0, 1]. The availability of photometric
redshift estimates is accounted for by integrating $\eta(r,z)$
as in (\ref{phot}), and finally we obtain each of the model
correlation functions on a grid. Note that the intrinsic models are 
calculated using the relationships for the distances $f(w)$ pertaining to 
the $\Lambda$CDM cosmology described above, and that we use 
$\xi_{\rm gg}(r)=(r/5\,h^{-1}\,{\rm Mpc})^{-1.8}$.

\subsubsection{Cross-term basis models}
A simple toy-model prescription for the cross-term is adopted, parameterising $\bar{\xi^{\rm LI}}$ directly, and evaluated for the $\Lambda$CDM cosmology. The main feature of the model is the dependence on angular diameter distance as in HS04; for fixed 
$z_{i}< z_{j}$, the cross-term at a certain angular scale increases with $z_{j}$. To describe the angular dependence, an exponential term is used so that the observed signal is
\begin{equation}
\bar {\xi^{\rm LI}}(\theta,{\bar z_{i}},{\bar z_{j}})= T\;{\rm exp}\,\left(-\theta/\theta_{\rm \times}\right)\,\int \frac{f_k(w_i,w)}{f_k(w)}p(z)dz\;,
\end{equation}
where $T$ is a normalisation and $\theta_{\times}$ is a scale angle. The normalisation of the templates is fixed so that the magnitude of the cross-term is 12\% of the $\Lambda$CDM lensing signal in the lowest angular separation bin, with respect to redshift slices at $z=0.62$ and $z=1.13$, and is negative as in HS04. Since the intrinsic alignment and cross-term models are currently poorly constrained, and this normalisation is model dependent, it has been chosen to result in a cross-term that is sub-dominant to the intrinsic signal for sources with small physical separations, and to dominate the intrinsic signal and have a measurable effect on the two point correlation function for sources with larger physical separations - in other words a cross-term that would have to be accounted for in the analysis of cosmic shear survey data. Note that for the templates, this normalisation is however arbitrary, since the signal is projected into the template functions, determining best-fit estimates of their amplitudes. 
Two sets of templates are used for the cross-term: a set of three templates with $\theta_{\times}$ set to [1, 5, 10] /3000 radians, and a set of nine templates with $\theta_{\times}$ values [0.05, 0.1, 0.3, 0.1, 1.0, 2.0, 4.0, 6.0, 8.0] /3000 radians (where 1/3000 radian $\approx$ 1.15$\arcmin \approx 0.55$ Mpc at $z=1$). Again, since predictions for the cross-term encompass a wide range of amplitudes, the template set would be expanded on as necessary.

\section{Results}
The results are presented in the context of possible future multi-colour cosmic shear surveys, with field size $L=14\arcmin$, hence the largest scale on which the ellipticity correlation function
is available is $\sqrt 2\times 14\arcmin$. A galaxy
number density of 30\,arcmin$^{-2}$ and ellipticity dispersion of
$\seps=0.3$ (typical of ground-based observations) are used throughout.  We consider surveys of various areas, the fiducial survey 
having ${\cal N_{F}}=300$ independent pointings. The value of $\sigma_{\rm phot}=0.1$ 
is chosen since this is characteristic of that obtained with current SED fitting procedures such as hyper-z using a wide range of optical and near-infrared filters (Bolzonella et al.\, 2000). There are $N_{Z}=65$ redshift slices between ${\bar z}=0.2$ and 2.12, and $N_{K}=25$ angular separation bins between $0\arcminf 3$ and $15\arcmin$.

The galaxy redshift distribution follows the parameterisation from 
Smail et al. (1995), i.e. $p({\bar z})=\beta/[z_{0}\,\Gamma_{\beta}(3/\beta)]({\bar z}/z_{0})^{2}\,{\rm exp}\left(-({\bar z}/z_{0})^{\beta}\right)$, where $\Gamma_{\beta}$ denotes the gamma function. We take $\beta=3/2$ and $z_{0}=2/3$ yielding ${\ave {\bar z}}\approx 1$.

\subsection{Minimising the contamination from the intrinsic and cross correlation signals}
We start with a survey of ${\cal N_{F}}=300$ fields. Our  ``observed" correlation functions
$\bar{\xi}^{\rm obs}_{m}$ comprise a lensing, an intrinsic and a cross 
contribution. The intrinsic alignment model for spirals from HRH, 
$\eta(r,z)=0.012\,{\rm exp}(-r/1.5\,h^{-1}{\rm Mpc})$ was used to obtain $\bar{\vec{\xi}^{\rm I}}$, the lensing signal $\bar{\vec{\xi}^{\rm L}}$ was calculated for a $\Lambda$CDM cosmology and $\bar{\vec{\xi}^{\rm LI}}$ for an exponential model with a scale angle of 4.8/3000 radians (which will be our fiducial cross-term model). In this subsection, the larger set of nine cross-term templates was 
used.

\begin{figure}
\begin{center}
\epsfig{file=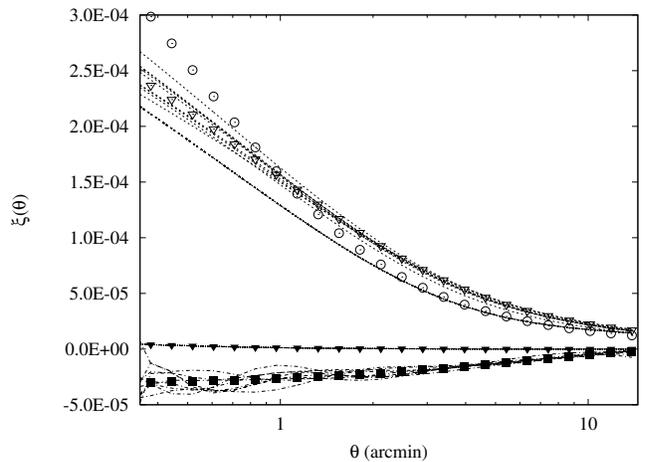,scale=0.7}
\caption{Contributions to the two point correlation function, as a function of angular scale, for slices at $z=0.62$ and $z=1.13$. The inverted open triangles show the lensing correlation function for $\Lambda$CDM (and for comparison open circles show OCDM). Squares (inverted filled triangles) show the cross-term (intrinsic term) 
as described in the text. The sets of 10 lines show recovered lensing, cross and intrinsic signals for different noise realisations; the recovered instrinsic signals are indistinguishable on the figure, since they are very similar to each other.} The total input signal is the curve offset below the recovered lensing 
signals. The survey has 300 fields each 14$\arcmin$ on a side. 
\label{plot1}
\end{center}
\end{figure}

\begin{figure}
\begin{center}
\epsfig{file=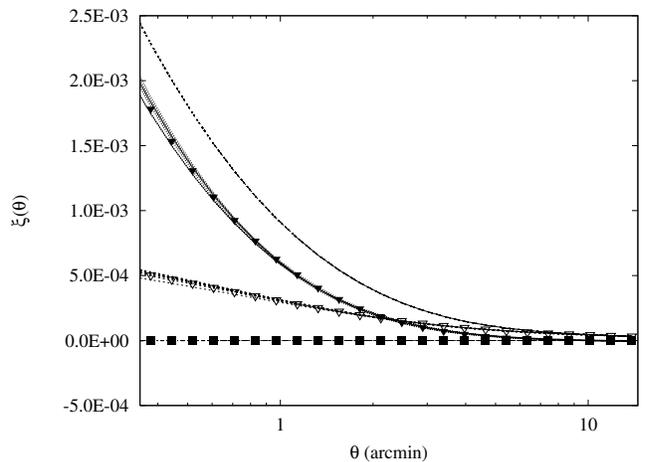,scale=0.7}
\caption{As for Fig.\,1 but for slices at $z=1.07$ and $z=1.13$. The lensing correlation function for 
OCDM is not displayed. The total input signal is the uppermost curve.}
\label{plot2}
\end{center}
\end{figure}

Random gaussian distributed errors with dispersion $\sigma={\cal C}_{mm}^{0.5}$ were added to these correlation functions giving noise realisations, and best-fit parameters ${\cal G}_{n}$ were recovered. Fig.\,\ref{plot1} and Fig.\,\ref{plot2} show the (noise-free) input and recovered correlation functions between two 
combinations of redshift slices: Fig.\,1 shows correlations for slices at ${\bar z}\sim 0.6$ and ${\bar z}\sim 1$ and Fig.\,2 shows the correlations between two neighbouring slices at ${\bar z}\sim 1$.
In both figures we show the total signal, and in Fig.\,\ref{plot1} we also plot the lensing correlation function expected for an OCDM cosmology. 

For bins with a large separation in redshift, the intrinsic signal is negligible, as expected. Similarly, for bins with a small separation in redshift, the cross-term is negligible.  The reduced $\chi^{2}$  values of the recovered fits to the noise realisations are $\approx 1$. Note that if we were simply to down-weight physically close pairs to minimise any intrinsic alignment signal, that neglecting the cross-term may cause e.g. an underestimation of $\sigma_8$.

\begin{figure}
\begin{center}
\epsfig{file=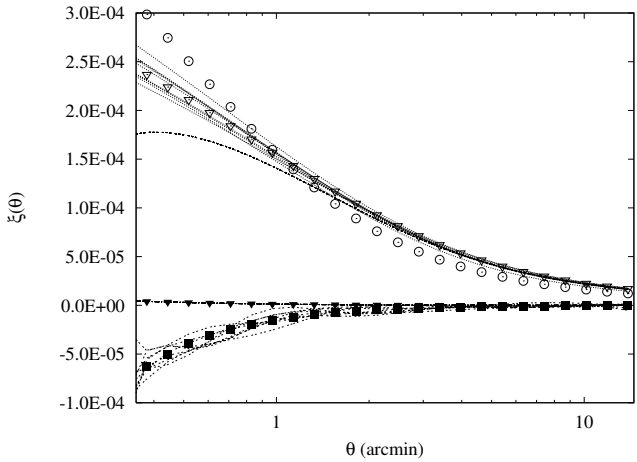,scale=0.7}
\caption{As for Fig.\,1 but for a cross-term with a power-law dependence on angular separation.}
\label{plot3}
\end{center}
\end{figure}

\begin{figure}
\begin{center}
\epsfig{file=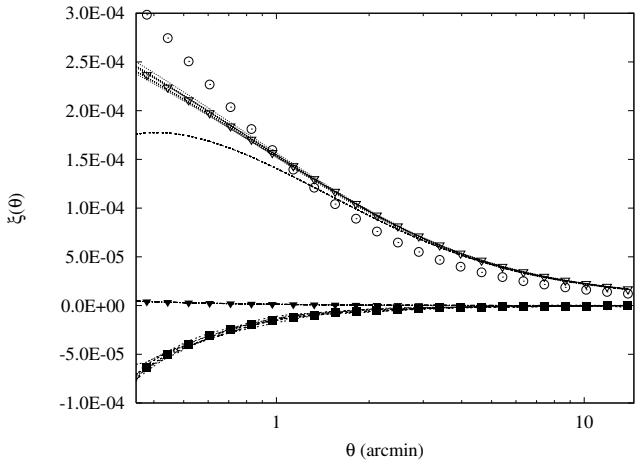, scale=0.7}
\caption{As for Fig.\,1 but for a cross-term with a power-law dependence on angular separation, and for a survey with 3000 fields.}
\label{plot4}
\end{center}
\end{figure}

This procedure was repeated, this time with the angular dependence of the cross-term taken to be a power-law $\propto \theta^{-1.5}$. Note that this functional form is not contained in the template set, providing a check of the robustness of the method. Fig.\,\ref{plot3} shows the (noise-free) input and recovered correlation functions between two 
redshift slices at ${\bar z}\sim 0.6$ and ${\bar z}\sim 1$, along with the total signal, and the OCDM lensing correlation function. We also show in Fig.\,\ref{plot4} the corresponding results for ${\cal N_{F}}=3000$ (i.e. a survey of $\approx$160 sq. degrees, similar to the ongoing CFHTLS). Note than in practice, the template sets for both the IA and cross-term would encompass a wide range of models; this is particularly important for the cross-term as discussed below.

\subsection{Lifting the $\Omega_{\rm m}-\sigma_8$ degeneracy}
A well-known degeneracy in cosmic shear constraints is between $\Omega_{\rm m}$ and 
$\sigma_{8}$. How accurately can a survey with ${\cal N}_{f}=3000$, 1000 or 300 distinguish
between a $\Lambda$CDM model and a near-degenerate model ($\Omega_{\rm m}=0.4, \sigma_{8}=0.78$) for the lensing model, in the presence of both an intrinsic alignment and a 
cross signal?

The gridded correlation functions for lensing in the $\Lambda$CDM and degenerate cosmologies 
were taken and added to the gridded HRH intrinsic correlation function, along with the fiducial cross-term model. The simulations involved generating 10000 (1000) noise realisations for the 3000 and 1000 (300) field surveys, to obtain the ``observed" gridded correlation functions ${\vec{\bar\xi}}_{\Lambda\rm CDM}$ and ${\vec{\bar\xi}}_{\rm degen}$. The best-fit amplitudes ${\hat{\cal G}}_{n}$ for the template functions were recovered. A set of thirteen template functions was used containing a) the nine models for intrinsic alignments, b) the set of three models for the cross-term, along with c) either the $\Lambda$CDM lensing template or the degenerate model template in turn. We end up with four sets of best-fit amplitudes for each size of survey, and corresponding values of $\chi^{2}$ for how well the signal is modeled by projection into the templates: two sets generated using the $\Lambda$CDM (degenerate) lensing template, one of which is projected into a template set containing the $\Lambda$CDM lensing template, and the other projected into a template set containing the degenerate lensing template. Values were determined for $\Delta\chi^{2}(\Lambda{\rm
CDM}-{\rm degen})$ and (ii) $\Delta\chi^{2}({\rm degen}-\Lambda{\rm
CDM})$, corresponding to the difference in goodness-of-fit for the
noise realisations when ${\vec{\bar\xi}}_{\Lambda{\rm CDM}}$ and
${\vec{\bar\xi}}_{\rm degen}$ are used in the template set. 
When the fiducial $\Lambda$CDM model (degenerate
model) is the best-fit and is contained in the template set, values of
$\Delta\chi^{2}$ should be negative. This gives a
measure of our ability to differentiate between models using
correlation functions between redshift slices.

For ${\cal N}_{f}=3000$, when the model for 
the cosmology is contained in the template set, in {\em all} cases the noisy correlation functions for that cosmology are better fit. When ${\cal N}_{f}=1000 (300)$, and ${\vec{\bar\xi}}_{\Lambda\rm CDM}$ is
contained in the template set, in 99.77 (93.9) \% of cases the noisy
$\Lambda$CDM correlation functions are better fit. Also, when
${\vec{\bar\xi}}_{\rm degen}$ is in the template set, 99.88 (94.4) \% of the
noisy degenerate correlation functions have better fits. Hence, within
the assumptions made, and in the presence of intrinsic and cross-term signals, 
these two cosmological models could be distinguished well in excess of the  
3-$\sigma$ level when ${\cal N}_{f}=1000$, and just in excess of 2-$\sigma$ 
when ${\cal N}_{f}=300$.

For comparison, simulations were undertaken without any intrinsic and cross-term signal, so the only signal was from cosmic shear in a $\Lambda\rm CDM$ cosmology for a survey with ${\cal N}_{f}=300$. When ${\vec{\bar\xi}}_{\Lambda\rm CDM}$ is the lensing template for recovery, in just over 99\% of cases the noisy correlation functions are better fit, compared with when ${\vec{\bar\xi}}_{\rm degen}$ is the lensing template. The same holds for simulations where the cosmic shear signal is from the degenerate cosmology, and ${\vec{\bar\xi}}_{\rm degen}$ is the lensing template.

Although in the foregoing example, the form of the cross-term is quite close to the templates, this is achievable in practice since the template set used would be large. Nevertheless, for the observed signal, let us next replace the cross-term model with exponential dependence on angular separation, with the power-law model (scale dependence $\propto \theta^{-1.5}$), and use the expanded set of nine exponential cross-term templates. The $\Lambda$CDM model for lensing, HRH instrinsic alignment model,and power-law model for the cross-term were used to generate 10000 noise realisations of a ${\cal N}_{f}=1000$ survey. Best-fit amplitudes for the template functions were determined in the same manner as above, first when the template set only contains a $\Lambda$CDM template, and second when it only contains the degenerate template. Again, the two cosmological models can be distinguished in excess of the 3-$\sigma$ level.

\section{Discussion and conclusions}
Recently, it was suggested by HS04 that the lensing correlation
function may be contaminated, not only by intrinsic galaxy alignments, but also by a 
cross-term which results in a correlation between the ellipticity of a background 
lensed galaxy and a foreground galaxy. The current and next generation of cosmic 
shear surveys have the potential to probe the evolution of the matter power spectrum, 
and to test other families of cosmological models (e.g. quintessence models) with similar physical manifestations as the concordance model  (e.g. Benabed \& Bernardeau 2001; Tereno et al. 2005). 
In this era of precision cosmology, it is therefore essential to have the ability to quantify and isolate 
the contribution from both intrinsic galaxy alignments and the instrinsic-lensing cross-term 
in order to remove these systematics. Although the intrinsic and intrinsic-lensing terms are important in their own right, providing constraints on physical processes at the epoch of galaxy formation, here we
have concentrated on their role as contaminants of the pure lensing signal.
 
That the lensing, intrinsic and cross-term have different dependency on redshift enables the 
components to be isolated when photometric redshifts are available. In particular, the impact of the potential cross-term suggested by HS04, which is impossible to remove by simply neglecting or down-weighting physically close galaxy pairs, can be  minimised using this type of tomographic procedure.
Since the dependence on redshift of the lensing signal is more akin to the cross-term than to 
the intrinsic signal, a wide range of template functions for each of the terms, coupled with a large survey area, is beneficial in obtaining an estimate of the cross-term magnitude. 

Application to a real survey will warrant as large a set of template functions as possible, particularly a wide range of functional forms for the intrinsic and cross-terms. One strategy for the lensing templates could be to have correlation functions corresponding to cosmological models close to our best estimate for the cosmological parameters - perhaps finely sampling models which are degenerate in the $\Omega_{\rm m} - \sigma_{8}$ plane.  
If the reduced $\chi^{2}$ of the best fit is significantly larger than 1, this indicates that additional template functions do need to be included. Additional constraints on intrinsic alignment templates could come from direct measurements of the pure intrinsic alignment signal in low redshift surveys where it could dominate lensing (e.g. Brown et al. 2002), or in surveys where spectroscopic redshifts are available.
For the cross-term, HS04 note that an estimate of the density-intrinsic shear correlation has already been made on small scales in the context of assessing the contamination of galaxy-galaxy lensing in the SDSS (Hirata et al. 2004b). Such measurements on larger scales would help to define the template set, albeit at a low redshift so that the possibility of evolution would have to be accounted for.

The main goal of the method developed here and in KS03 is to harness redshift information in order to minimise the contamination of the lensing signal by any intrinsic alignment or lensing-intrinsic signal. Although a comparison between how well discrete cosmological models fit the data can be made (such as in the discussion above of models degenerate in the $\Omega_{\rm m}-\sigma_{8}$ plane),  the intention is not to determine best-fit cosmological parameters. In the absence of an intrinsic alignment or cross-term, Simon, King \& Schneider (2004) have demonstrated how even crude redshift binning can greatly decrease errors on cosmological parameters determined by measuring shear correlation functions. Almost an order of magnitude improvement is obtained when a few redshift bins are employed, and progressively little is gained as the number of bins is increased. For cosmological parameter constraint, a standard technique with much cruder redshift binning, and a much finer sampling of cosmological parameters could be applied to the cleaned lensing signal obtained from the method presented here. Such techniques are typically analogous to the use of eq.\,(8), but in the form
\begin{equation}
S=(\vec{\bar{\xi}}^{\rm obs}-\vec{\xi(\pi)})\,{\cal C}^{-1}\,(\vec{\bar{\xi}}^{\rm
obs}-{\vec{\xi(\pi)}})\;,
\end{equation} 
where $\vec{\xi(\pi)}$ is the theoretical shear correlation function for a set of cosmological parameters $\pi$. 

The simulations in this paper have been carried out for a field of 14$\arcmin$ on a side; this is 
similar to the imager on VLT-VIMOS. Having a larger field would increase the number of pairs of galaxies in a particular bin from which the correlation function is measured, so decreasing the shot 
noise. The accuracy with which photometric redshifts can be determined for the galaxies in a survey also impacts on the results: upcoming multi-colour surveys such as the DES (Dark Energy Survey) will have redshift  accuracies of $\delta z = 0.03-0.07$ for galaxies out to $z\sim 1.1$, depending on redshift and galaxy type (white paper at http://decam.fnal.gov). Cosmic shear surveys are carried out at optical frequencies, but when wide-field infrared observations of the same fields are carried out (with telescopes such as VISTA), this will greatly assist photometric redshift estimation. 

With our restricted set of exponential templates for the cross-term, and an exponential form for its true amplitude, a survey of $\sim 60$ sq. degrees, with photometric redshifts for the galaxies used in the analysis, would enable us to distinguish between models degenerate in $\Omega_{\rm m} - \sigma_{8}$ in excess of the 3-$\sigma$ level. When the angular dependence of the cross-term is a power-law form, the same degree of distinction between models requires an expanded set of (exponential) template functions. Keeping in mind our simplified ansatz for the covariance matrix, this would crudely correspond to $\sim 100$ sq. degrees in practice - although given such a large number of fields we beat down cosmic variance. Accurate representations of the covariance matrix will be calculated using ray-tracing through large N-body simulations. 

\begin{acknowledgements}
I thank George Efstathiou, Max Pettini, Peter Schneider and Uro\u{s} Seljak for discussions, and the referee for a very helpful and constructive report. This work was supported by the Royal Society. 
\end{acknowledgements}

\def\ref#1{\bibitem[1998]{}#1}

\end{document}